\documentclass{aa}  

\usepackage{graphicx}
\usepackage{txfonts}
\usepackage{lipsum}
\usepackage{subcaption}         % necessary for continued figures, example in section 3
\usepackage{lscape}             % to rotate a single page table, example in appendix.
\usepackage{placeins}           % useful with \FloatBarrier, to keep 
\usepackage{color} %\definecolor{purple}{rgb}{0.5,0,0.87}

\newcommand\dm{{\rm DM}}
\newcommand\sidm{{\rm SIDM}}
\newcommand\ufd{{\rm UFD}}
\newcommand\nfw{{\rm NFW}}
\newcommand\kms{${\rm km\,s^{-1}}$}
\newcommand\cmg{${\rm cm^2\,g^{-1}}$}

\begin{document}
\title{
  Constraints on dark matter models from the stellar cores observed in ultra-faint dwarf galaxies:
}
\subtitle{Self-interacting dark matter}

\author{Jorge S\'anchez Almeida\inst{1,2}}
\institute{
       {Instituto de Astrof\'\i sica de Canarias, La Laguna, Tenerife, E-38200, Spain}\\
       \email{jos@iac.es}
       \and
       Departamento de Astrof\'\i sica, Universidad de La Laguna
       }

   \date{Received \today; accepted \dots}

% \abstract{}{}{}{}{}
% 5 {} token are mandatory

   \abstract{
  It has been proposed that the stellar cores observed in ultra-faint dwarf (\ufd) galaxies reflect underlying dark matter (\dm) cores that cannot be formed by stellar feedback acting on collisionless cold dark matter (C\dm) halos. Assuming this claim is correct, we investigate the constraints that arise if such cores are produced by self-interacting dark matter (\sidm).
We derive the range of \sidm\ cross-sections ($\sigma/m$) required to reproduce the observed core sizes. These can result from halos in either the core-formation phase (low $\sigma/m$) or the core-collapse phase (high $\sigma/m$), yielding a wide allowed range ($\sim$0.3\,--\,200\,\cmg) consistent with values reported in the literature for more massive galaxies.
We also construct a simple model relating stellar mass to core radius -- two observables likely connected in \sidm . This model reproduces the stellar core sizes and masses in \ufd s with $\sigma/m$ consistent with those derived above. It also predicts a trend of increasing core radius with stellar mass, in agreement with observations of more massive dwarf galaxies.
The model central \dm\ densities match observations when assuming the \sidm\ profile to originate from an initial C\dm\ halo that follows the mass–concentration relation.
Since stellar feedback is insufficient to form cores in these galaxies, \ufd s unbiasedly anchor $\sigma/m$ at low velocities.
If the core-collapse scenario holds  (i.e., high $\sigma/m$), \ufd\ halos are thermalized on kpc scales, approximately two orders of magnitude larger than the stellar cores. These large thermalization scales could potentially influence substructure formation in more massive systems.
%As a byproduct, we provide a stellar-to-halo mass relation spanning from $10^8$ to $10^{15}\,M_\odot$, and show that the observed stellar distributions in \ufd s are well matched by the cored \sidm\ halo profile adopted in the simple model.
}

\keywords{
  Galaxies: dwarf --
  Galaxies: evolution --
  Galaxies: halos --
  Galaxies: kinematics and dynamics --
  Galaxies: stellar content --
  dark matter
}

               \maketitle

\titlerunning{}
               \authorrunning{J. S\'anchez Almeida}

%%%%%%%%%%%%%%%%%%%%%%%%%%%%%%%%%%%%%%%%%%%%%%%%%%%%%%%%%%%%%%
\begin{nolinenumbers}
%%%%%%%%%%%%%%%%%%%%%%%%%%%%%%%%%%%%%%%%%%%%%%%%%%%%%%%%%%%%%%
\section{Introduction}\label{sec:intro}

The nature of dark matter (\dm) remains one of the most pressing open questions in  science\footnote{{\tt https://www.nature.com/collections/mnwshvsswk}}. In the standard cosmological model, dark matter (\dm) is assumed to be cold and collisionless (CDM), consisting of particles that interact with baryons and with themselves solely through gravity. While this model has proven remarkably successful, it is likely an effective approximation of a deeper, more complex theory in which \dm\ exhibits some form of non-gravitational interaction. To advance our understanding, it is therefore crucial to explore potential tensions between CDM predictions and observational data \citep[see this argument in, e.g.,][]{2017nuco.confa0101S,2021arXiv210602672P,2025RSPTA.38340022E}. 

One notable tension that extends from the well-known core-cusp problem involves the dark matter distribution in the faintest galaxies. In cosmological simulations, self-gravitating C\dm\ halos develop steep central density profiles, or cusps, which contrast with the approximately constant-density cores observed in many galaxies \citep[e.g.,][]{2017Galax...5...17D,2017ARA&A..55..343B,2019A&ARv..27....2S}. According to the standard interpretation, the C\dm\ cusps are transformed into cores through baryonic processes -- such as feedback from star formation -- that redistribute matter and alter the total mass profile, including the dark matter component \citep{2010Natur.463..203G,2012MNRAS.421.3464P}. However, the effectiveness of this mechanism depends on the availability of energy from star formation, and it becomes negligible below a stellar mass threshold of roughly $10^6\,M_\odot$ \citep[e.g.,][]{2012ApJ...759L..42P,2016MNRAS.456.3542T}. Ultra-faint dwarf galaxies (\ufd s), with stellar masses well below this limit \citep[e.g.,][]{2019ARA&A..57..375S,2022NatAs...6..659B}, are therefore expected to retain cuspy dark matter halos if the \dm\ were C\dm. Thus, the detection of \dm\ cores in such systems, as recently reported by \citet[][]{2024ApJ...973L..15S}, deepens the core-cusp problem and points toward physics beyond the standard CDM paradigm.

Determining the \dm\ distribution in the tiny \ufd s is challenging and remains subject to significant uncertainty. Obtaining detailed kinematic data is difficult, so the recent claim relies instead on the fact that their stellar distributions exhibit conspicuous cores. Under a set of defensible assumptions -- such as non-tangentially biased velocities and axial symmetry -- the Eddington Inversion Method allows one to demonstrate that a stellar core is physically inconsistent with a cuspy \dm\ profile \citep[][]{2006ApJ...642..752A,2023ApJ...954..153S,2024A&A...690A.151S}. This approach, implemented in the tool described by \citet{2025A&A...694A.283S}, was applied to the surface density profiles of the six \ufd s observed by \citet{2024ApJ...967...72R}, leading to the conclusion that these galaxies likely reside in cored \dm\ potentials. Once alternative explanations -- such as stellar feedback -- are ruled out, the most compelling interpretation is that the \dm\ departs from the standard C\dm\ behavior.

Many physical models beyond C\dm\ produce \dm\ cores (e.g.,
  self-interacting  \dm , \citeauthor{2000PhRvL..84.3760S}~\citeyear{2000PhRvL..84.3760S};
fuzzy \dm , \citeauthor{2014NatPh..10..496S}~\citeyear{2014NatPh..10..496S};
self-interacting fuzzy \dm , \citeauthor{2025arXiv250204838I}~ \citeyear{2025arXiv250204838I};
warm \dm, \citeauthor{2001ApJ...556...93B}~\citeyear{2001ApJ...556...93B};
Bose-Einstein condensate \dm , \citeauthor{2023MNRAS.518.4064D}~\citeyear{2023MNRAS.518.4064D};
fermionic \dm, \citeauthor{2023Univ....9..197A}~\citeyear{2023Univ....9..197A};
or late-time \dm\ decay, \citeauthor{2018JCAP...07..013C}~\citeyear{2018JCAP...07..013C}). 
The ease with which cores form is understandable, given that cores are characteristic of self-gravitating systems approaching thermodynamic equilibrium \citep{1993PhLA..174..384P,2008gady.book.....B,2022Univ....8..214S}. Any physical process leading to the thermalization of the \dm\ halo would develop a central core.
In fact, what is truly remarkable about the CDM framework is that, due to the absence of non-gravitational interactions, halos fail to reach thermodynamic equilibrium within a Hubble time ($t_H$) and instead retain memory of their initial conditions \citep[e.g.,][]{2020MNRAS.495.4994B}.

Although the formation of cores is a generic prediction, the constraints arising from the presence of \dm\ cores in \ufd\ depend on the specific physical \dm\ model and therefore needs to be analyzed individually for each case. Here, we focus on the constraints applicable to self-interacting dark matter (\sidm ) models, in which \dm\ particles undergo collisions beyond standard two-body gravitational interactions. This new interaction is characterized by a cross section sufficiently large to ensure that the \dm\ halos of \ufd\ galaxies thermalize within $t_H$. \sidm\ represents the most straightforward conceptual extension of the standard CDM framework, introducing particle collisions that promote thermalization—much like molecular interactions establish thermodynamic equilibrium in the air.
The concept was first introduced by \citet{2000PhRvL..84.3760S} and has since gained considerable attention in the literature \citep[e.g.,][]{2001sddm.symp..263W,2015MNRAS.453...29E,2018PhR...730....1T,2022arXiv220710638A}, partly because self-interactions are a generic prediction of many particle physics \dm\ models involving a hidden sector \citep[e.g.,][]{2009JCAP...07..004F,2010PhRvD..81h3522B}, and partly because of the profound impact self-interactions would have ruling out many popular dark matter particle candidates \citep[e.g.,][]{2018NatAs...2..856S}.

The paper is organized as follows:
Sect.~\ref{sec:sidmxsec} uses the scaling relation provided by \citet{2023MNRAS.523.4786O} to work out the \sidm\ cross-sections able to generate the cores observed in \ufd s. They are compared with \sidm\ cross-section estimates from the literature, derived using a variety of methods including gravitational lensing, galaxy cluster collisions, galaxy rotation curves, and dynamical modeling of dwarf galaxies.
Section~\ref{sec:toy_model_new} uses the time evolution from cuspy to cored \dm\  profiles, as modeled by \citet{2024JCAP...02..032Y}, to construct a toy model that relates the stellar core radius to the stellar mass of a galaxy for a given velocity-dependent \sidm\ cross-section. According to this model,  the stellar cores in \ufd s require cross-sections similar to the ones inferred in Sect.~\ref{sec:sidmxsec}. Moreover, it shows the trend of increasing stellar core radius with stellar mass followed by other more massive dwarfs from the literature.
Cross-sections consistent with the observed \ufd\ cores can be small, if the \dm\ halos are in the phase of forming a core, or large, in they are in the core-collapse phase. Even though both solutions are possible, the physical mechanism giving rise to the \dm\ distribution in the halo is different. Section~\ref{sec:thermalr} shows how only the \dm\ particles forming the core have interacted during the core-formation phase whereas the rest maintain the original distribution expected from collision-less  C\dm\ halos. In the core-collapse phase, however, the whole halo is in thermodynamic equilibrium so that stellar cores are tiny compared with the thermalization radii.  
Section~\ref{sec:conclusions} discusses and summarizes the results of the work.
Appendixes~\ref{sec:appa} and  \ref{sec:appc} detail the literature on cross-sections and other properties of the galaxies used for contextualization, whereas  Appendix~\ref{sec:appb} gives the stellar mass versus \dm\ halo mass relation employed in the toy model. Appendix~\ref{app:sidm_fit} shows how the stellar distribution observed in \ufd s is well reproduced by the halo shape employed in Sect.~\ref{sec:toy_model_new} to construct the toy model.  

%
%%%%%%%%%%%%%%%%%%%%%%%%%%%%%%%%%%%%%%
% 
\begin{figure*}
\centering
\includegraphics[width=0.6\linewidth]{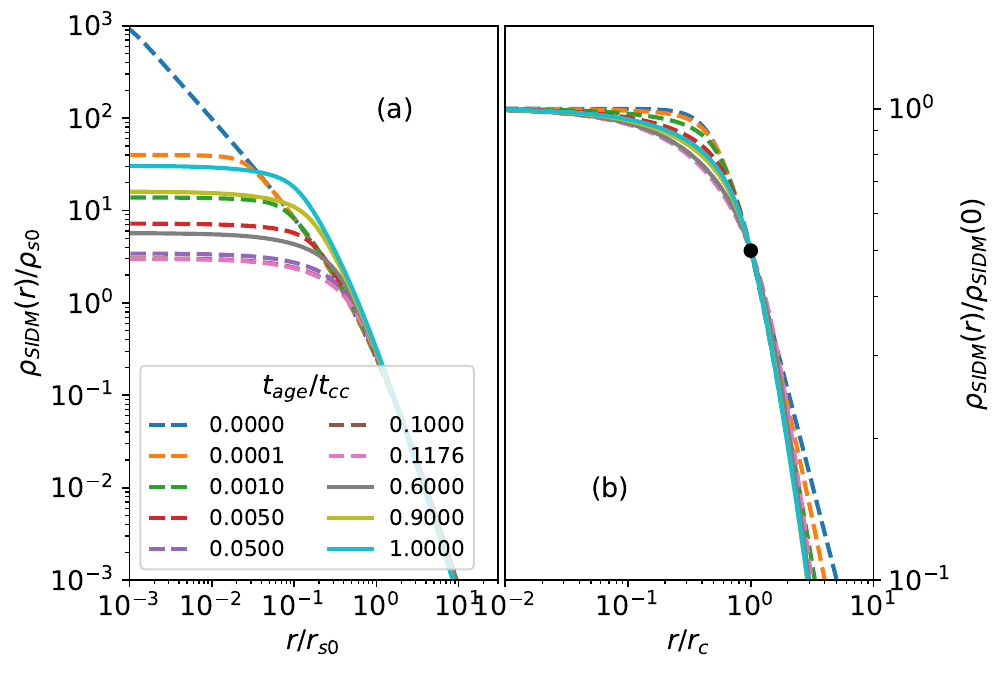}
\includegraphics[width=0.6\linewidth]{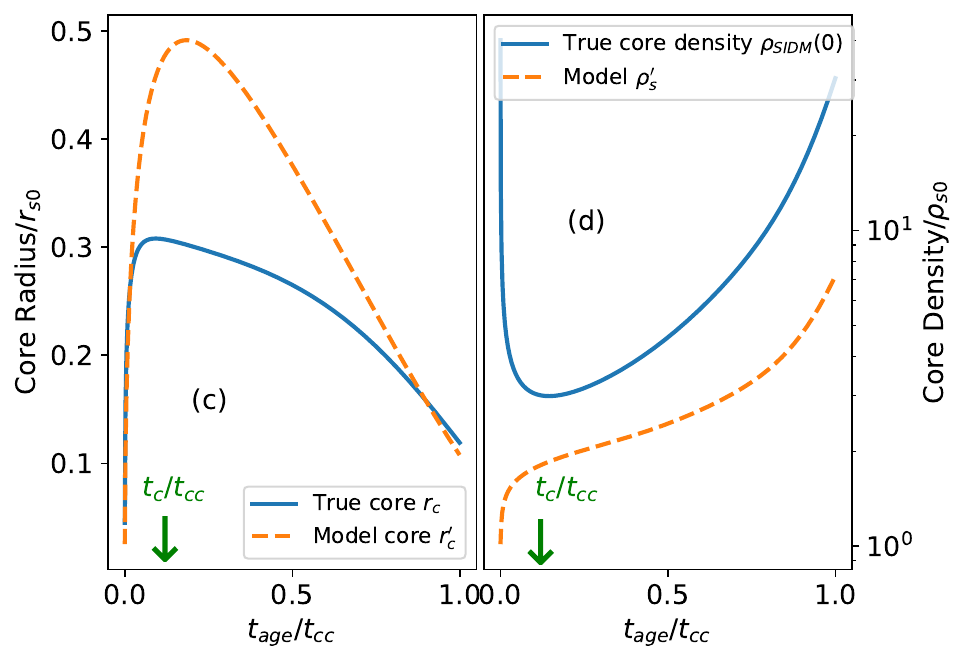}
\caption{Time evolution of \sidm\ halos according to \citet{2024JCAP...02..032Y}. The time scale is parameterized in terms of the core-collapse timescale, $t_{cc}$ -- Eqs.~(\ref{eq:master_outm}) and (\ref{eq:cctime}). The halos start off as \nfw\ profiles of characteristic density and radius $\rho_{s0}$ and $r_{s0}$, respectively.
  (a)~Radial density at different timesteps going from almost the initial conditions ($t_{age}/t_{cc}= 0$) to core-collapse  ($t_{age}/t_{cc}= 1$), including the formation of a maximum core  ($t_{age}/t_{cc}= t_c/t_{cc}\simeq 0.12$). Profiles before the maximum core formation ($t_{age} \leq t_c$) are represented as dashed lines whereas profiles after this core-formation are shown as solid lines.  (b)~Same profiles as (a) but normalized to the central density and to the core radius defined in Eq.~(\ref{foot:1}). The black bullet symbol indicates the location of the core radius. 
(c)~Time variation of the true core (the blue solid line; $r_c$ in Eq.~[\ref{eq:rcore2}]) and of the model core (the orange dashed line; $r'_c$ in Eq.~[\ref{eq:rho_sim}]). 
(d)~Time variation of the central density (the blue solid line; $\rho_\sidm(0)$) and of the characteristic density (the orange dashed line; $\rho_s'$ in Eq.~[\ref{eq:rho_sim}]). The green arrow in (c) and (d) marks the time scale for maximum core formation.
}
\label{fig:dmcoresize11}
\end{figure*}
\section{Relating core radius with \sidm\ cross section}\label{sec:sidmxsec}
The constraints on the cross sections can be estimated keeping in mind that SIDM halos of all masses and sizes seem to follow the same time evolution once properly normalized to the characteristic timescale for the collision of two \dm\  particles  \citep{2002ApJ...568..475B,2011PhRvL.106q1302L},
  \begin{equation}
    \tau_{coll} = \frac{1}{\sigma\,v\,\rho/m},
    \label{eq:tcoll}
\end{equation}
where $\sigma$, $v$, $\rho$, and $m$ stand for the cross-section, the relative velocities between \dm\ particles, the characteristic volume density, and the \dm\ particle mass, respectively. The universal behavior has been shown to hold under various assumptions including constant and velocity dependent cross section \citep{2002ApJ...568..475B,2023MNRAS.523.4786O,2024JCAP...02..032Y}. Figure~\ref{fig:dmcoresize11} illustrates this universal time evolution as computed by \citeauthor{2024JCAP...02..032Y}~(\citeyear{2024JCAP...02..032Y}; Eq.~[3.3]) from the N-body numerical simulation by \citet{2022JCAP...09..077Y} with $\sigma=\sigma(v)$. The original NFW halos \citep[][]{1997ApJ...490..493N} turn to form a core that expands while dropping the central density to a minimum value (Fig.~\ref{fig:dmcoresize11}a, the dashed lines, and Figs.~\ref{fig:dmcoresize11}c and \ref{fig:dmcoresize11}d, the blue solid lines). From this time on, the \dm --\dm\ collisions trigger the core-collapse of the \dm\ halo that begins to contract increasing its central density while loosing part of the core mass \citep[e.g.,][]{2015MNRAS.453...29E,2023MNRAS.523.4786O}; see Fig.~\ref{fig:dmcoresize11}a, the dashed lines.
The scaling between $\sigma$ and the true time scale driving the evolution of the system has to be found through numerical simulations. In the scanning of initial conditions computed by \cite{2023MNRAS.523.4786O}, they begin from DM halos with NFW profiles characterized by density $\rho_s$ and length scale $r_s$,
  \begin{equation}
    \rho_\nfw(r) =\frac{\rho_s}{(r/r_s) (1+r/r_s)^2},
    \label{eq:nfw}
\end{equation}
which evolve with time due to the \dm\ self-collisions treated as heat conduction. Although initial \dm\ halos may deviate from NFW profiles with somewhat steeper inner power-law slopes \citep[e.g.,][]{2004ApJ...607..125T,2023MNRAS.518.3509D}, the evolution of SIDM halos is likely insensitive to these initial conditions because DM particle self-interactions thermalize the distribution, leaving halo properties determined mainly by global quantities such as mass and energy \citep[e.g.,][and references therein]{2022Univ....8..214S}. The characteristic timescale to reach the smallest density and the largest core radius (hereinafter, core-formation timescale) is found to be,
\begin{equation}
t_{c} \simeq 6.75~{\rm Gyr} \frac{0.6}{C}
\frac{10\,{\rm cm^2 g^{-1}}}{\sigma/m}
\frac{100\,{\rm km\,s^{-1}}}{v_{max}}
\frac{10^{-2} M_\odot\,{\rm pc^{-3}}}{\rho_s},
\label{eq:master_outm}
\end{equation}
with the symbol $C$ a numerical factor of order one introduced by the conductivity model representing the collisions. On the other hand, $v_{max}$ is the maximum of the rotational velocity of the original NFW halo and so depends on $r_s$ and $\rho_s$ as 
\begin{equation}
  v_{max}\simeq 1.65\, r_s\sqrt{G\,\rho_s},
  \label{eq:vmax}
\end{equation}
with $G$ the gravitational constant. The \sidm\ cross-section in Eq.~~(\ref{eq:master_outm}) is an effective cross-section that integrates over the halo characteristic velocities and provides a good approximation to self-scattering during the halo evolution \citep[see ][]{2022JCAP...09..077Y,2023MNRAS.523.4786O}.
To fit our needs, Eq.~(\ref{eq:master_outm}) can be rewritten  as
\begin{equation}
  t_{c} \simeq 2.5\,{\rm Gyr}\,
  \frac{0.6}{C}\,
  \frac{10\,{\rm cm^2 g^{-1}}}{\sigma/m}\
  \left[\frac{44\,M_\odot\,{\rm pc^{-2}}}{\rho_c(t_c)\,r_c(t_c)}\right]^{3/2}
  \left[\frac{r_c(t_c)}{50\,{\rm pc}}\right]^{1/2},
  \label{eq:master_outm2}
\end{equation}
where $r_c(t_c)$ is the largest core radius which, for the model analyzed by \cite{2023MNRAS.523.4786O} is just $r_c(t_c)\simeq 0.45\, r_s$.
Here and throughout, the core radius ($r_c$) is implicitly defined as 
\begin{equation}
\rho(r_c)=\rho(0)/2,
\label{foot:1}
\end{equation}
with $\rho(r)$ the density profile. The symbol  $\rho_c(t_c)$ stands for the smallest core density, reached when the radius is largest,  which happen to be $\rho_c(t_c)\simeq 2.4\,\rho_s$. The normalization of $\rho_c(t_c)\,r_c(t_c)$ to $44\,M_\odot\,{\rm pc^{-2}}$ was set because this product, for a large range of \dm\ halo masses, has been observed to be approximately constant \citep{1995ApJ...447L..25B,2009MNRAS.397.1169D} with the value used in Eq.~(\ref{eq:master_outm2}) \citep{2025Galax..13....6S}.
%
%(\cite{2024JCAP...02..032Y} also work out $t_c$, and they find an expression identical to Eq.~(\ref{eq:master_outm2}) except for a multiplicative factor of order one.)
%
After the formation of the core,  the halo evolves to eventually core collapse \citep[e.g.,][]{2008gady.book.....B}.  The central density increases while the core shrinks to a point where they become similar to the original NFW cusp (Figs.~\ref{fig:dmcoresize11}c and ~\ref{fig:dmcoresize11}d). According to Fig.~\ref{fig:dmcoresize11}c and  the  numerical simulations used to derive  Eq.~(\ref{eq:master_outm}), this core-collapse occurs in a time scale of the order of
\begin{equation}
  t_{cc} \simeq 8.5~ t_c.
  \label{eq:cctime}
\end{equation}

The above equations allow us to estimate the cross-section required to produce the cores observed in \ufd s. Their halo must have developed a core but has not yet undergone full core-collapse.  If  $t_{age}$ is the age of such halo, then  the condition imposes 
\begin{equation}
  \frac{1}{2}t_{c}  \leq \beta_{age}\,t_{H} \leq t_{cc},
\label{eq:inequality}
\end{equation}
with  $\beta_{age}=t_{age}/t_{H}$ the age relative to the Hubble time. The lower limit in Eq.~(\ref{eq:inequality}) comes from the assumption that cores are already formed at a time $t_c/2$, which is consistent with the simulations (see Fig.~\ref{fig:dmcoresize11}a and Sect.~\ref{sec:toy_model_new}).   
Equations~(\ref{eq:master_outm2}), (\ref{eq:cctime}), and (\ref{eq:inequality}) can be combined to yield
\begin{equation}
\frac{1}{2}\,\frac{\sigma_H/m}{\beta_{age}}   \leq \sigma/m\leq 8.5\,\frac{\sigma_H/m}{\beta_{age}},
\label{eq:sigma_range}
\end{equation}
where the new symbol $\sigma_H/m$ stands for
\begin{equation}
  \sigma_H/m = 1.85\,{\rm cm^2 g^{-1}}\,\frac{0.6}{C}\, \left[\frac{44\,M_\odot\,{\rm pc^{-2}}}{\rho_c(t_c)\,r_c(t_c)}\right]^{3/2} \left[\frac{r_c(t_c)}{50\,{\rm pc}}\right]^{1/2},
\label{eq:sigmaH}
\end{equation}
which is the cross-section required to form the largest core of radius $r_c$ with density $\rho_c$ in a Hubble time $t_H$.
\begin{figure}
  \centering
\includegraphics[width=1\columnwidth]{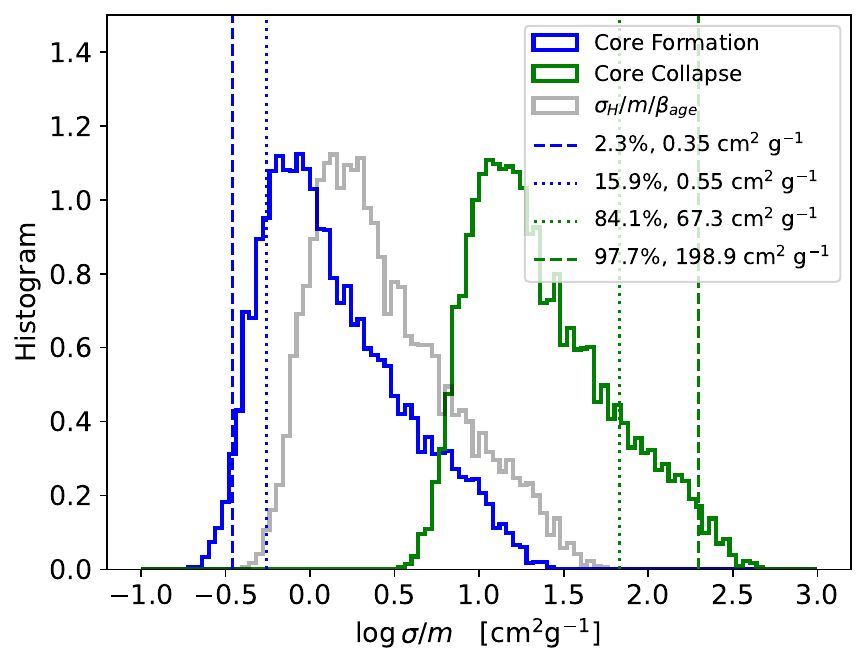}
\caption{Monte-Carlo simulation used to estimate the \sidm\ cross-sections required to account for the \dm\ cores of UFDs. The cross-sections depend on a number of poorly constrained parameters  (the parameters in Eq.~[\ref{eq:sigmaH}]). The Monte-Carlo simulations allows us to propagate their uncertainties on $\sigma_H/m/\beta_{age}$ (the gray histogram) and, via Eq.~(\ref{eq:sigma_range}), on the cross-section $\sigma/m$. The blue histogram represents the distribution assuming the UFDs are in the process of forming the core whereas the green histogram assumes that \dm\ es evolving to core-collapse. The vertical dashed lines mark the 2.3\% and 97.7\% percentiles used to constraint the range of viable cross-sections given in Eq.~(\ref{eq:ufdlimit}). See the main text for further details.  
}
\label{fig:dmcoresize14}
\end{figure}
Figure~\ref{fig:dmcoresize14} shows the range of cross-sections permitted by Eq.~(\ref{eq:sigma_range}) given the range of properties of the \ufd s analyzed by \citet{2024ApJ...973L..15S}.
The distribution of the parameter $\sigma_H/m/\beta_{age}$ (gray histogram in Fig.~\ref{fig:dmcoresize14}) was obtained using a Monte Carlo simulation in which all unknown parameters were varied independently, each drawn from a uniform probability distribution within physically reasonable ranges. In particular,
$C$ was varied between 0.5 and 0.8, consistent with values from numerical simulations \citep[e.g.,][]{2023MNRAS.523.4786O},
$\beta_{age}$ between 0.3 and 0.7, in line with dark matter halo ages reported in simulations \citep[e.g.,][]{2006MNRAS.368.1931L,2015MNRAS.452.1217C},
$\rho_c(t_c)r_c(t_c)$ between 10 and 100~$M_\odot\,{\rm pc}^{-2}$, as inferred from observations \citep[e.g.,][]{2025Galax..13....6S},
and $r_c(t_c)$ between 20 and 60~pc, following the observational constraints of \citet{2024ApJ...973L..15S}.
According to Eq.~(\ref{eq:sigma_range}), the cores can be produced with cross-sections going from  $(\sigma_H/m/\beta_{age})/2$, if the halos are forming the core (the blue histogram in Fig.~\ref{fig:dmcoresize14}), to $(\sigma_H/m/\beta_{age})\times 8.5$, if the halos are approaching core-collapse (the green histogram).
Since we do not know whether the \ufd s are forming their cores or core-collapsing, we consider the two possibilities to set limits on their cross-section. Thus, $\sigma/m$ has to be in the range 
\begin{equation}
  0.35\,{\rm cm^2\,g^{-1}} < \sigma/m < 199\,{\rm cm^2\,g^{-1}},
  \label{eq:ufdlimit}
\end{equation}
which corresponds to the $-2$\,SD (Standard Deviation)  limit of the $(\sigma_H/m/\beta_{age})/2$ distribution (percentile 2.3\%) and the $+2$\,SD limit of the  $(\sigma_H/m/\beta_{age})\times 8.5$ distribution (percentile 97.7\%; see the dashed lines in Fig.~\ref{fig:dmcoresize14}). The interval in Eq.~(\ref{eq:ufdlimit}) can be written as
\begin{equation}
  \sigma/m= 10^{0.92\pm 1.37}\,{\rm cm^2\,g^{-1}}.
 % \log\left[\frac{\sigma/m}{1\,{\rm cm^2\,g^{-1}}}\right]= 0.91\pm 1.37.
%  \log(\sigma/m)= 0.91\pm 1.37,
\label{eq:ufdlimit3}
\end{equation}

In order to assign a  velocity to the UFDs ($v_{eff}$) suitable to characterize the cross-section, we follow \citet{2024JCAP...02..032Y} where
\begin{equation}
  v_{eff} = 0.64\,v_{max},
\label{eq:veff}
\end{equation}
with $v_{max}$ given by Eq.~(\ref{eq:vmax}). Consistently with the analysis by \citet{2023MNRAS.523.4786O} that leads to Eq.~(\ref{eq:master_outm}), $r_c(t_c)\simeq 0.45\, r_s$ and $\rho_c(t_c)\simeq 2.4\,\rho_s$, therefore,  assuming  $\rho_c(t_c)\,r_c(t_c)$ and $r_c(t_c)$, one has $v_{max}$ (Eq.~[\ref{eq:vmax}]) and so $v_{eff}$  (Eq.~[\ref{eq:veff}]). The same Monte-Carlo simulation providing the interval in Eq.~(\ref{eq:ufdlimit}) renders
\begin{equation}
v_{eff} = 4.6^{+2.7}_{-2.5}\,{\rm km\,s^{-1}}.
\label{eq:veff2}
\end{equation}

\subsection{Comparison with existing estimates of $\sigma/m$}\label{sec:comparison}
\begin{figure*}
  \centering
\includegraphics[width = 0.8\linewidth]{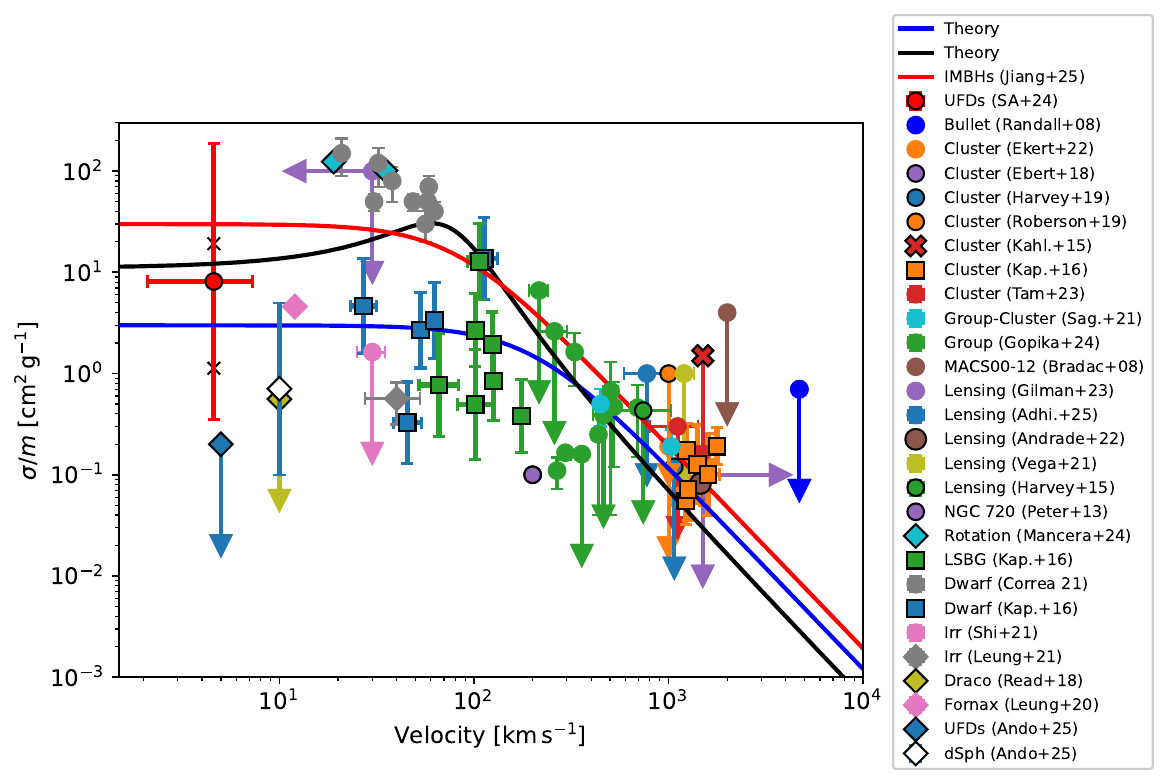}
\caption{
  Summary of the \sidm\ cross-sections found in the literature. They are plotted versus a  representative relative velocity of the \dm\ particles in the observed object.  The references are labeled in the inset with the equivalence with actual papers given in Appendix~\ref{sec:appa}. This appendix also points out how $\sigma/m$ and   $Velocity$ mean something  different in the different works, which may explains part of the scatter. The value provided by the \ufd s analyzed in this work is the red circle with error bars at the lowest velocities, with the two times symbols representing the median of the core formation and the core-collapse distributions shown in Fig.~\ref{fig:dmcoresize14}.
The solid lines corresponds to the analytical form in Eq.~(\ref{eq:resonance}) where $\sigma_0/m, w_0, w_1= 30$~\cmg , 0, 80~\kms\  (the red line), 3~\cmg , 0, 200~\kms\ (the blue line),  and  11~\cmg , 60~\kms , 45~\kms (the black line). }
\label{fig:read_Ghosh+22Fig9}
\end{figure*}
Figure~\ref{fig:read_Ghosh+22Fig9} provides a summary of the current constraints and claims on the \sidm\ cross-section found in the literature. The sources are quite diverse and there is no general rule on how $\sigma/m$ and $Velocity$ are computed from the observational data. We select these cross-sections from the papers citing the seminal work where \citet{2008ApJ...679.1173R} set an upper limit to $\sigma/m$ from the lack of separation between \dm\ and stars in the Bullet Cluster. The individual references are pointed out in Appendix~\ref{sec:appa} together with some general remarks on how each cross-section was derived.

The constraints provided by the \ufd s are incorporated into Fig.~\ref{fig:read_Ghosh+22Fig9} as a single point,
 with error bars representing the range of possible cross-sections. Specifically, the red circle with error bars at the lowest velocity corresponds to the cross-section range given in Eq.~(\ref{eq:ufdlimit3}), and to the velocity range defined by $v_{eff}$ in Eq.~(\ref{eq:veff2}). The two times symbols along the vertical error bar mark the median values of $\sigma/m$ obtained from the core-formation and core-collapse distributions shown in Fig.~\ref{fig:dmcoresize14}. Several key conclusions can be drawn from the figure: (1) \ufd s align well with the overall trend. (2) Since stellar feedback cannot account for the presence of cores in these systems, \ufd s provide a clean  anchor point for constraining the velocity-dependent cross-section at low velocities. This is in contrast to more luminous dwarf galaxies, where core formation may be significantly influenced by baryonic feedback (see Sect.~\ref{sec:intro}). (3) The observed scatter in cross-section values across different galaxies may arise from known but uncorrected biases (such as neglecting baryonic effects) as well as from less well-understood systematics, including variations in the definitions of velocity and cross-section. Alternatively, it may reflect intrinsic diversity in galaxy properties, as proposed by, e.g., \citet{PhysRevLett.119.111102}.

%%%%%%%%%%%%%%%%%%%%%%%%%%%%%%%%%%%%%%%%%%%%%%%%%%%%%%%%%%%%%%
\begin{figure*}
\centering
\includegraphics[width=0.9\linewidth]{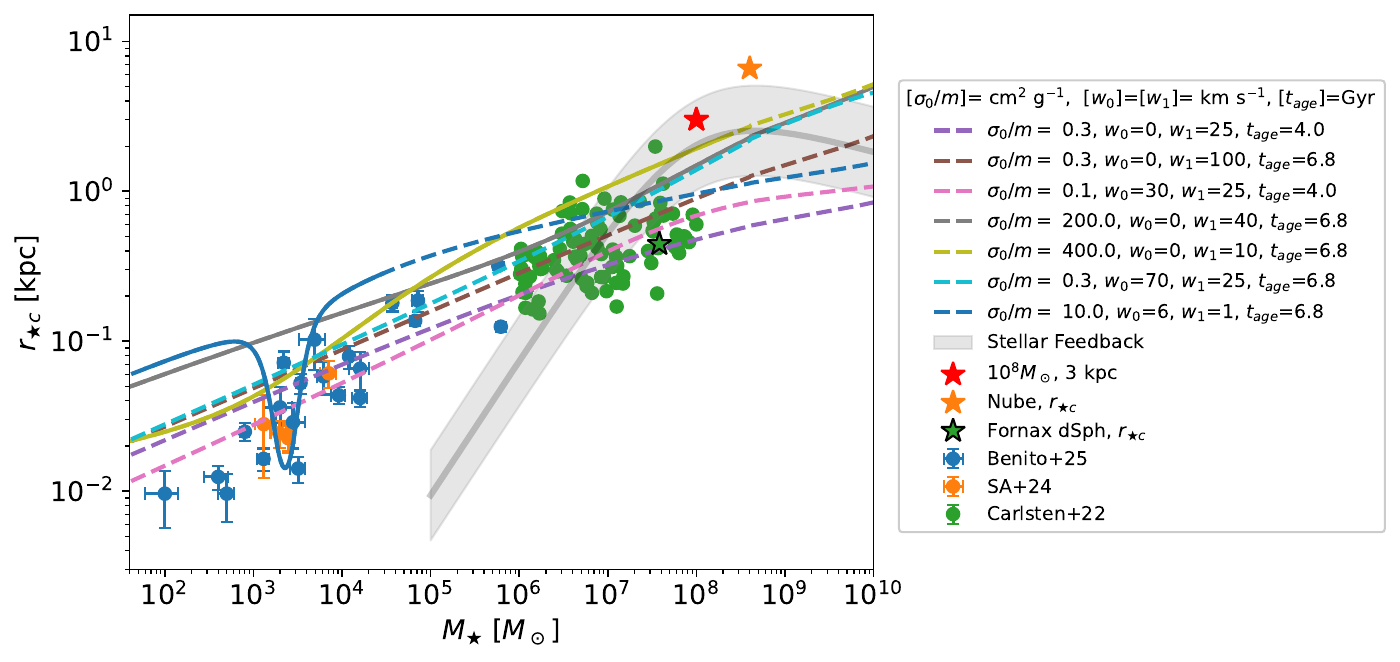}
\caption{Stellar core radius versus stellar mass as predicted by the simple model presented in Sect.~\ref{sec:toy_model_new}.  Each line corresponds to the variation of $r_{\star c}$ with $M_\star$ when the \dm\ halo mass goes  from $10^7$ to $10^{14}\,M_\odot$. The type of line depends on whether the corresponding profile is in the core-formation phase ($t_{age} < t_c$; dashed line) or in the core-collapse phase ($t_{age} > t_c$; solid line). The parameters that define the cross-sections are given in the inset, with the actual velocity dependence represented in Fig.~\ref{fig:dmcoresize13_3} with the same color code employed here. Observations of $r_{\star c}$ versus $M_\star$ are represented as symbols, each one corresponding to an individual object. The measurements are labelled in the inset, with the \ufd s analyzed in this work appearing under the label {\tt SA+24}. The link between labels and references is specified in Appendix~\ref{sec:appc}. The figure also includes the theoretical expectations from stellar feedback on C\dm\ halos  from \citet[][the gray band and solid line]{2020MNRAS.497.2393L}.}
\label{fig:dmcoresize13}
\end{figure*}
\begin{figure}
\centering
\includegraphics[width=1.0\linewidth]{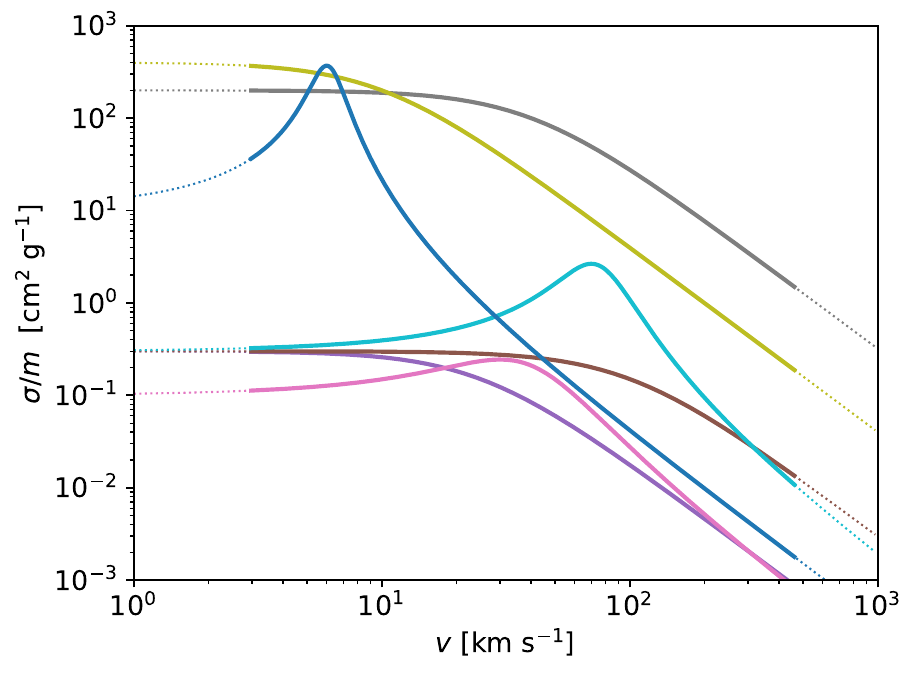}
\caption{Cross sections used to compute $r_{\star c}(M_\star)$ in Fig.~\ref{fig:dmcoresize13}. The color code is the same in both figures. The range of velocities highlighted with thicker lines corresponds to the range of effective velocities (Eq.~[\ref{eq:veff}]) when the halo masses go from $10^7$ to $10^{14}\,M_\odot$, as used in Fig.~\ref{fig:dmcoresize13}.
}
\label{fig:dmcoresize13_3}
\end{figure}
\begin{figure}
\centering
\includegraphics[width=1.0\linewidth]{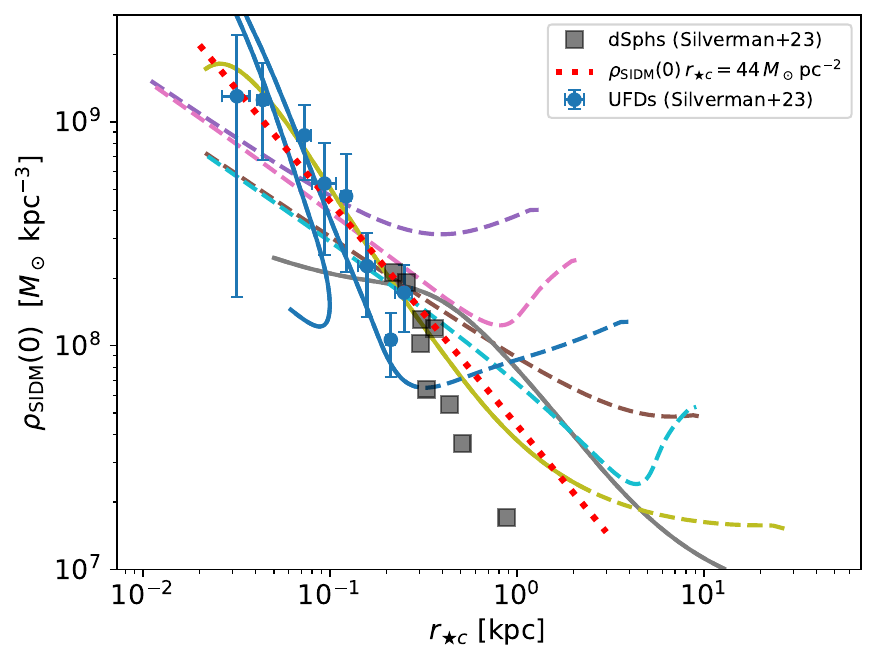}
\caption{Central density of the \sidm\ profiles used in Fig.~\ref{fig:dmcoresize13}. It is represented versus the stellar core-radius with the same color code employed in Figs.~\ref{fig:dmcoresize13} and \ref{fig:dmcoresize13_3}. As in Fig.~\ref{fig:dmcoresize13}, the line is dashed or solid depending on whether the corresponding profile is in the core-formation phase (dashed line) or in the core-collapse phase (solid line).
The figure also includes a number of observed densities for \ufd s and dSph collected by \citet[][their Fig.~10]{2023MNRAS.518.2418S}. The range of central densities observed in \ufd s is well covered by the \sidm\ simulations, both in the core-formation and the core-collapse phase.     
The increase of central density with decreasing radius fits in well with the observational fact that the product central density times core radius tends to be constant independent of the \dm\ halo mass, with a value around $44\,M_\odot\,{\rm pc}^{-2}$. This trend is represented as the red dotted line, assuming $r_{c}\simeq r_{\star c}$. 
}
\label{fig:dmcoresize13_2}
\end{figure}
\section{Relation between stellar core radius and stellar mass set by \sidm}\label{sec:toy_model_new}
Two simple direct observables of a galaxy, stellar core radius and stellar mass, have to be related if self-interactions between \dm\ particles  are responsible for the core. The relation arises because the halo mass determines the \dm\ densities and relative velocities, which, together with the cross-section, govern the degree of halo evolution and thus its core radius (Fig. \ref{fig:dmcoresize11}c). With some reasonable assumptions, one can map \dm\ radius and mass to stellar radius and mass.  Here we work out a toy model for such a relation under a number of simplifying  assumptions as explained below. The relation between these two observables will be used to constraint the \sidm\ cross-section, complementing and checking for consistency  the exercise carried out in Sect.~\ref{sec:sidmxsec}.

As we explain in Sect.~\ref{sec:sidmxsec} and show in Fig.~\ref{fig:dmcoresize11},
\citet{2024JCAP...02..032Y} provide an analytical approximation for the time evolution of \dm\ halos from initial \nfw\ profiles to cored-collapsed halos. The halos are defined as
\begin{equation}
  \rho_\sidm(r) = \frac{1}{[(r/r_s')^4+(r_c'/r_s')^4]^{1/4}}\frac{\rho'_s}{(1+r/r_s')^2},
\label{eq:rho_sim}
\end{equation}
where the free parameters $\rho'_s$, $r'_s$, and $r'_c$ depend on the age of the halo relative to the core-collapse time scale (Eq.~[\ref{eq:cctime}]). Except when $r'_c= 0$,  this density presents an inner core of density
\begin{equation}
  \rho_\sidm(0)=\rho'_s\,r'_s/r'_c.
\label{eq:rho00}
\end{equation}
When $r'_c\to 0$,  Eq.~(\ref{eq:rho_sim}) describes a \nfw\ profile (Eq.~[\ref{eq:nfw}]), which is what happens as $t_{age}\to 0$ since $r'_c\to 0$ as well. We note that $r_c'$ parameterizes the core properties but is not the core radius that we employ ($r_c$ in Eq.~[\ref{foot:1}]), which for the \sidm\ profile is implicitly defined from the expression
\begin{equation}
  2 r'_c/r'_s = (1+r_c/r_s')^2\,[(r_c/r'_s)^4+(r'_c/r'_s)^4]^{1/4}.
\label{eq:rcore2}
\end{equation}
The parameters  $\rho'_s$, $r'_s$, and $r'_c$ are worked out in \citet{2024JCAP...02..032Y}, normalized to the value that $\rho_s'$ and  $r'_s$ have at $t_{age}=0$, i.e., when $\rho_\sidm$ was a $\rho_\nfw$ profile of constants $\rho_s$ and $r_s$ (Eq.~[\ref{eq:nfw}]). Examples are given in Fig.~\ref{fig:dmcoresize11}a, where profiles before the core-formation time ($t_{age} < t_c$) are shown as dashed lines whereas profiles after core formation are shown using solid lines. Figure~\ref{fig:dmcoresize11}b shows the same profile as  Fig.~\ref{fig:dmcoresize11}a  with different normalization -- they are normalized to the central density, Eq.~(\ref{eq:rho00}), and to the core radius, Eq.~(\ref{eq:rcore2}). Figures~\ref{fig:dmcoresize11}c and \ref{fig:dmcoresize11}c show how these two normalizing parameters vary with the age of the \dm\ halo. 
%
%\comment{for a footnote: Once can easily prove that Eqs.~(\ref{eq:master_outm}) and (\ref{eq:cctime}) provide the same core-collapse timescale as Eq.~(3.2) in  \citet{2024JCAP...02..032Y}.} % see dmcoresize12.py

In order to obtain the core radius corresponding to a \sidm\ density profile with a given halo mass ($M_h$) and age, one has to assume the properties of the initial \nfw\ halo and the \sidm\ cross-section. For the initial \nfw\ halo, we assume the halo mass concentration relation by \citet[][]{2025MNRAS.536..728S}, but this specific selection lacks significant impact on the results, as we analyze below.
As for the cross-section,  it varies with velocity $v$ as 
\begin{equation}
\frac{\sigma}{m}(v)=\frac{\sigma_0}{m}\,\frac{1+w_0^2/w_1^2}{1+\left[v-w_0\right]^2/w_1^2},
\label{eq:resonance}
\end{equation}
which describes the existence of a resonance of width $w_1$ at $v = w_0$, as expected for some types of \sidm\ particles \citep[see ][]{2013PhRvL.110k1301T,2019PhRvL.122g1103C,2020JCAP...06..043C}. More importantly, when $w_0\to 0$ the resonance goes away and Eq.~(\ref{eq:resonance}) describes the angular average cross-section commonly used in the literature and appearing  when the interaction between particles occurs through a Yukawa potential  \citep{2010PhLB..692...70I,2022MNRAS.517.3045C,2023ApJ...953..169T}. The constant $\sigma_0/m$ parameterizes the cross-section when $v\to 0$.
Examples of $\sigma/m(v)$ from Eq.~(\ref{eq:resonance}) are shown as the solid lines in Fig.~\ref{fig:read_Ghosh+22Fig9}, with the actual $\sigma_0/m$, $w_0$, and $w_1$ given in the caption.  
The above assumptions allow the determination of the \dm\ core radius as a function of halo mass, $r_c(M_h)$. In order to transform $r_c(M_h)$ to the observational plane stellar core radius ($r_{\star c}$) versus stellar mass ($M_\star$), two additional assumptions are needed, namely,
\begin{equation}
  r_{\star c}=\delta\ r_c,
  \label{eq:adhoc1}
\end{equation}
and
\begin{equation}
M_\star = M_\star(M_h).
\label{eq:adhoc2}
\end{equation}
We consider  $\delta\sim 1$ in Eq.~(\ref{eq:adhoc1})  as an ansatz  justified with the observation of some \ufd s \citep{2024ApJ...973L..15S}, but it can be relaxed easily, as we explore below. The mapping in Eq.~(\ref{eq:adhoc2}) is given in Appendix~\ref{sec:appb}, and it uses a recipe based on abundance matching for $M_h \gtrsim 10^{10}\,M_\odot$ \citep[][]{2013ApJ...770...57B} and on numerical simulations for $10^8\,M_\odot < M_h \lesssim 10^{10}\,M_\odot$  \citep[][]{2024arXiv240815214K}. Its limitations are also explored below. 

The stellar core versus stellar mass relation predicted by Eqs.~(\ref{eq:rho_sim})\,--\,(\ref{eq:adhoc2}) is shown in Fig.~\ref{fig:dmcoresize13}. Each line corresponds to the variation of $r_{\star c}$ with $M_\star$ when the \dm\ halo mass goes  from $10^7$ to $10^{14}\,M_\odot$ given a particular set of parameters. The type of line depends on whether the corresponding profile is in the core-formation phase ($t_{age} < t_c$; dashed line) or in the core-collapse phase ($t_{age} > t_c$; solid line) -- note that the type of line sometimes changes as $M_\star$ varies. The parameters that define the cross-sections are given in the inset ($\sigma_0/m$, $w_0$, and $w_1$; see Eq.~[\ref{eq:resonance}]), with the actual velocity dependence represented in Fig.~\ref{fig:dmcoresize13_3} employing the same the color code used Fig.~\ref{fig:dmcoresize13}. The inset also includes $t_{age}$, set to two values, $t_H/2$ and $t_H/3$. %
The blue line has been included to illustrate the impact on the core size of  the existence of a resonance at $\sim 6$\,km\,s$^{-1}$ (Fig.~\ref{fig:dmcoresize13_3});  it induces a sudden drop in $r_{\star c}$ at around $10^3\,M_\odot$ (Fig.~\ref{fig:dmcoresize13}).
The figure also includes a number measurements of $r_{\star c}$ versus $M_\star$ (the symbols) with  the \ufd s analyzed in this work appearing under the label {\tt SA+24} and represented with orange symbols. Figures~(\ref{fig:dmcoresize13}) and  (\ref{fig:dmcoresize13_3}) clearly shows that these \ufd\ observations are reproduced when $\sigma/m~\sim 0.1$\cmg , if the galaxies were in the core-formation phase (the purple or magenta lines) or $\sigma/m~\sim 300\,$\cmg, if the galaxies were in the core-collapse phase (the lemon yellow or blue lines). These fact is in fair agreement with the range worked out in Sect.~\ref{sec:sidmxsec} and given in Eqs.~(\ref{eq:veff}) and (\ref{eq:veff2}).
In addition to the \ufd s,  Fig.~(\ref{fig:dmcoresize13}) displays a number of measurements represented by symbols. These measurements are labelled in the inset, with the equivalence between labels and references specified in Appendix~\ref{sec:appc}.

Figure~\ref{fig:dmcoresize13} also includes the relation expected from stellar-feedback operating on C\dm\ halos,  as modeled by \citet[][the gray shaded area representing the scatter with the mean given by the solid line]{2020MNRAS.497.2393L}. Note the large discrepancy with observations at $M_\star < 10^6\,M_\odot$, which just illustrates the result that stellar feedback does not produce cores in low mass C\dm\ halos (see the arguments in Sect.~\ref{sec:intro}).  To transform the actual predictions to the observational plane, Eq.~(\ref{eq:adhoc1}) with $\delta=1$ has been assumed. We note that the stellar feedback sub-grid physic implemented in the numerical simulation used by  \citet[][]{2020MNRAS.497.2393L}  \citep[FIRE2;][]{2018MNRAS.480..800H} is particularly effective at transferring supernova kinetic  energy into the medium \citep[e.g.,][]{2020MNRAS.494.3971M,2021MNRAS.508.2979P} so that other alternative simulations will likely give even smaller cores, enhancing the tension with observations.
When  $M_\star > 10^6\,M_\odot$,  the figure suggests that stellar feedback has to be taken into account even if \sidm\ contributes to the formation of cores, since stellar feedback is present in real galaxies and it can potentially produce cores comparable in size to the observed ones. 

Figure \ref{fig:dmcoresize13_2} provides the central density of the \sidm\ profiles used in Fig.~\ref{fig:dmcoresize13}. They are represented versus  stellar core radius with the same color code employed in Figs.~\ref{fig:dmcoresize13} and \ref{fig:dmcoresize13_3}.  The figure also shows a number of observed densities for \ufd s and dSph as collected by \citet[][their Fig.~10]{2023MNRAS.518.2418S}. The range of central densities observed in \ufd s is well covered by the \sidm\ simulations, both in the core-formation and the core-collapse phase.  The increase of central density with decreasing radius by \citet[][]{2023MNRAS.518.2418S} fits in well the observational fact that the
product $\rho_\dm (0)\, r_c$ tends to be constant independently of the \dm\ halo mass, with a value around $44\,M_\odot\,{\rm pc}^{-2}$ \citep[][]{1995ApJ...447L..25B,2000ApJ...537L...9S,2016ApJ...817...84K,2025Galax..13....6S}. This value is represented as the red dotted line in Fig.~\ref{fig:dmcoresize13_2}, assuming $r_{\star c}\simeq r_{c}$.
We note that the relation $\rho_\mathrm{DM}(0)\, r_c \simeq \text{const}$ arises from the halo mass–concentration relation found in C\dm\ numerical simulations, under the assumption that the observed cores result from the redistribution of mass from an initial \nfw\ profile \citep[see][and references therein]{2025Galax..13....6S}. Consequently, any core density profile that originates from a \nfw\ profile consistent with the mass–concentration relation and reproduces the observed core size will naturally match the observed central dark matter density. The models employed both above and in Sect.~\ref{sec:sidmxsec} are of this kind, which explains why they also successfully recover the central densities.

\begin{figure}
\centering
\includegraphics[width=1.0\linewidth]{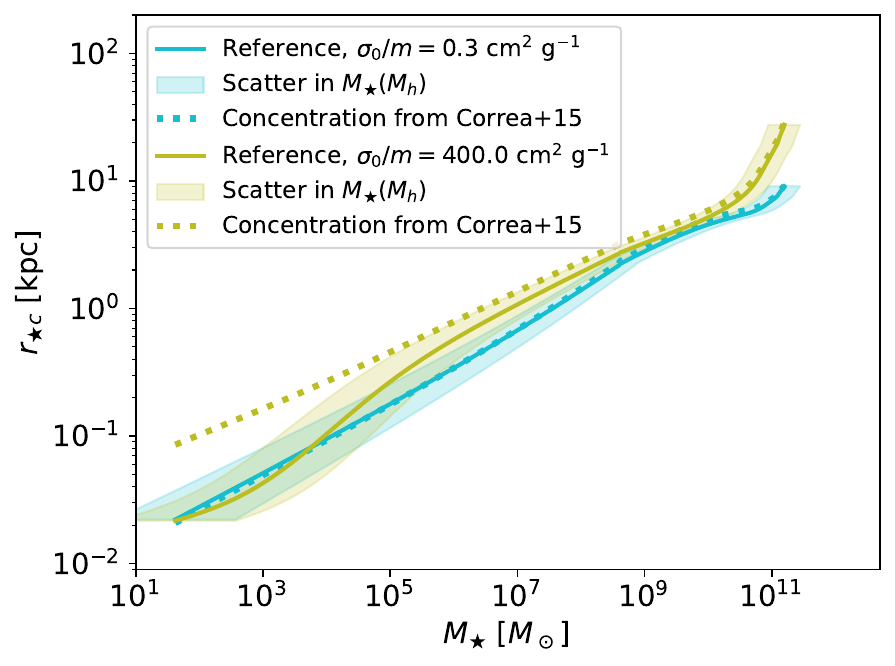}
\caption{
  Effect of the uncertainty in various parameters defining the toy model in Sect.~\ref{sec:toy_model_new}. We use two reference models, one with high cross-section (lemon orange) and one with low cross-section (cyan). They are represented  in Figs.~\ref{fig:dmcoresize13}\,--\,\ref{fig:dmcoresize13_2} with the same color.  As indicated in the inset, these reference models are drawn with solid lines whereas the shaded regions show the uncertainties in the stellar mass to halo mass relation. The dotted lines employ  a concentration to halo mass relation alternative to the one in the toy model. See main text for further details.
}
\label{fig:dmcoresize16}
\end{figure}
Figure~\ref{fig:dmcoresize16} explores the impact of the uncertainties in the free parameters that define the toy model. We modify the parameters for two models shown in Figs.~\ref{fig:dmcoresize13}\,--\,\ref{fig:dmcoresize13_2}. One of them has large cross-section (lemon yellow line) while the other has low cross-section (cyan line). They are represented in  Fig.~\ref{fig:dmcoresize16} with the same color as  in Figs.~\ref{fig:dmcoresize13}\,--\,\ref{fig:dmcoresize13_2}. The $M_\star$ versus $M_h$ relation (Eq.~[\ref{eq:adhoc2}]) has a scatter worked out in Appendix~\ref{sec:appb}, and this scatter gives rise to the colored bands in Fig.~\ref{fig:dmcoresize16} when the extreme $M_\star(M_h)$ relations are used in the toy model. The impact of the concentration to halo mass relation is analyzed by considering another relation, by \citet[][the dotted lines]{2015MNRAS.452.1217C}, often used in the literature. The uncertainty in the scaling between stellar  and \dm\ core radii ($\delta$ in Eq.~[\ref{eq:adhoc1}]) just shifts down all the curves according to the value of $\delta$. Neither this effect nor the change of age are included in Fig.~\ref{fig:dmcoresize16}.  Changing the age is formally identical to changing the amplitude of the cross-section since $t_{age}/t_{cc}$ depends on the product $t_{age}\,\sigma_0/m$ (Eqs.~[\ref{eq:master_outm}] and [\ref{eq:cctime}]). The effect is already illustrated in Fig.~\ref{fig:dmcoresize13}.  All in all, the differences between $r_{\star c}$ in the reference models and in all these alternatives are only around a factor of two, which is smaller than the effect of changing the parameters that define the cross-section ($\sigma_0/m$, $w_0$, and $w_1$ in Eq.~[\ref{eq:resonance}]) and smaller than the scatter among the observed galaxies (symbols in Fig.~\ref{fig:dmcoresize13}). 

A few conclusions that can be drawn from the  inspection of Figs.~\ref{fig:dmcoresize13}\,--\,\ref{fig:dmcoresize16}:
(1) The cross-section required to explain the \ufd s analyzed in Sect.~\ref{sec:sidmxsec}  (the orange symbols) are in a range broadly consistent with Eqs.~(\ref{eq:ufdlimit}) and (\ref{eq:veff2}).
(2) This toy model predicts an overall increase of the core size with increasing mass, which is in qualitative agreement with observations. Other models of \dm , like Fuzzy \dm , do not comply with this observational constraint \citep[e.g.,][]{2020ApJ...904..161B,2025PDU....4902010B}. Stellar feedback qualitatively predicts the trend, but quantitatively falls too short to explain the stellar core sizes at $M_\star< 10^6\,M_\odot$  (the gray shaded region in Fig.~\ref{fig:dmcoresize13}). 
(3) The uncertainty in the free parameters do not change the global trend of the dependence. Just produce an uncertainty in $r_{\star c}$ within a factor of two, which propagates into the value of $\sigma_0/m$ required to explain the observations.  
(4) The observed scatter is too large for the toy model, even when the uncertainties are considered. The origin of this {\em diversity} remains to be explained but it may be due to the difference pathways followed by the different galaxies as advocated by \citet{PhysRevLett.119.111102} or \citet{2025PhRvD.111j3041R}.    

%
%%%%%%%%%%%%%%%%%%%%%%%%%%%
%
\begin{figure}
\centering
\includegraphics[width=1.0\linewidth]{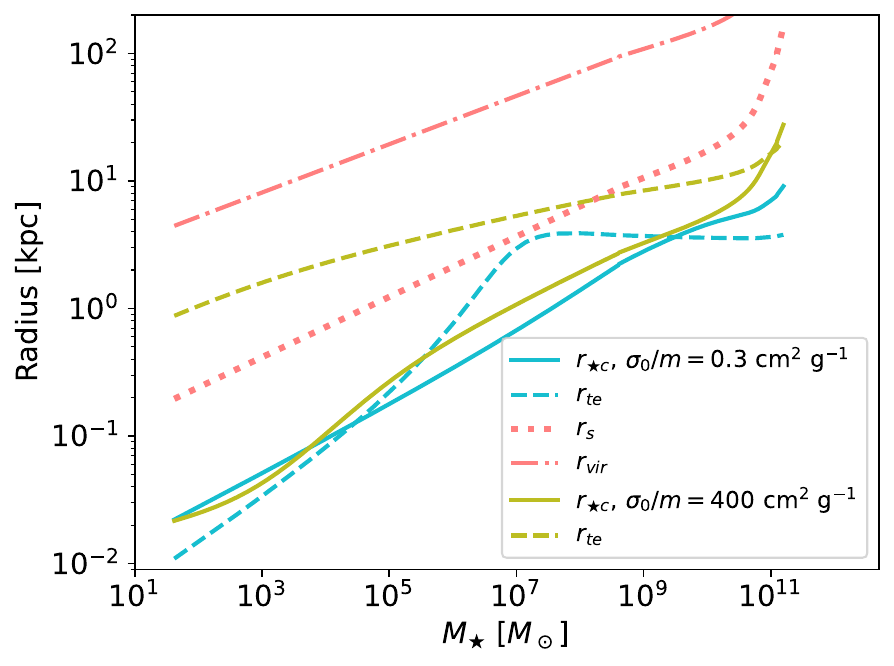}
\caption{
Comparison between the thermalization radius $r_{te}$ (Eq.~[\ref{eq:tage}]) and the stellar core radius $r_{\star c}$. When the cross-section is small ($\sigma_0/m\,=\,0.3$\,\cmg ; the cyan lines) then $r_{te}$ and $r_c$ are similar, in particular, $r_{te}\simeq r_{\star c}$ in the region of interest for \ufd s where $M_\star < 10^6\,M_\odot$. On the contrary,  $r_{te}\gg r_{\star c}$ for large cross-sections ($\sigma_0/m\,=\,400$\,\cmg ; the lemon yellow lines). For reference, the characteristic radius of the original \nfw~($r_s$; the red dotted line) and the virial radius ($r_{vir}$; the red dotted-dashed line) are included. 
}
\label{fig:dmcoresize17}
\end{figure}
\section{Thermalization radius}\label{sec:thermalr}
Sections~\ref{sec:sidmxsec} and \ref{sec:toy_model_new} show how the cores of \ufd s can be produced by a small \sidm\ cross-section, if the halos are in the core-formation phase,  or by a large cross-sections, if the halos are in the core-collapse phase (Fig.~\ref{fig:dmcoresize14}). Although the overall halo shape appears similar in both phases \citep[see  Fig.~\ref{fig:dmcoresize11}b and][]{2024JCAP...02..032Y}, the underlying physical processes differ. In the core-formation phase, only the core region is thermalized in the sense that only the \dm\ particles within this inner core region have had enough time to collide and thermalize. Thus, the halo outskirts still reflect the \nfw\ shape from which they originated. On the other hand, in the core-collapse phase, a significant part of the whole halo is thermalized, and its structure reflexes the distribution expected in self-gravitating \dm\  halos in thermodynamic equilibrium (see Sect.~\ref{sec:intro}).

In oder to work out the region of the halo that is thermalized in the two cases, we define the thermalization radius, $r_{te}$, as the radius where
\begin{equation}
  t_{age}\simeq \frac{1}{\rho_\nfw(r_{te})\,v_\nfw(r_{te})}\,\frac{1}{[\sigma/m]\left(v_\nfw(r_{te})\right)},
  \label{eq:tage}
\end{equation}
which follows from Eq.~(\ref{eq:tcoll}) with $\rho_\nfw(r)$ the \nfw\ profile defined in Eq.~(\ref{eq:nfw}) and $v_\nfw(r)$ a representative relative velocity between particles in the halo, taken to be the velocity dispersion. Assuming isotropy for the velocity field, it is $\sqrt{3}$ times the radial velocity dispersion, which is approximately  the circular velocity in \nfw\ halos \citep[e.g.,][]{2001MNRAS.321..155L}. Thus,
\begin{equation}
  v_\nfw^2(r)\simeq 3\,GM_\nfw(<r)/r,
\label{eq:velocity}
\end{equation}
with
\begin{equation}
  M_\nfw(<r)= 4\pi\rho_s r_s^3\,\left[\ln(1+r/r_s)-\frac{r/r_s}{1+r/r_s}\right].
  \label{eq:mass}
\end{equation}
We note that the product $\rho_\nfw(r)\,v_\nfw(r)$ is a monotonically decreasing function of $r$ so that, according to Eq.~(\ref{eq:tage}), the halos are expected to be thermalized inside-out as $t_{age}$ increases.  Moreover, since  $\rho_\nfw(r)\,v_\nfw(r)\propto r^{-1/2}$ when $r\to 0$, Eq.~(\ref{eq:tage}) always has solution for any $t_{age}$ provided $\sigma/m$.   % dmcoresize7.py

The cross-section to be used in our estimate are the same used everywhere else in the work (Eq.~[\ref{eq:resonance}]). Therefore, once $\sigma_0/m$, $w_0$, $w_1$ and mass of the \dm\ halo are set,  the thread of assumptions and arguments made  in Sect.~\ref{sec:toy_model_new} allows us to compute $r_{te}=r_{te}(M_\star)$.  This function is shown in Fig.~\ref{fig:dmcoresize17} for the two extreme cross-sections employed in Fig.~\ref{fig:dmcoresize16}; $\sigma_0/m\,=\,0.3$\,\cmg\ and 400\,\cmg . When the cross-section is small (cyan lines), the thermalization radius (dashed line) and the stellar core radius derived in Sect.~\ref{sec:toy_model_new} (solid line) are similar, in particular, $r_{te}\sim r_{\star c}$ in the region of interest for \ufd s where $M_\star < 10^6\,M_\odot$. On the contrary, for large cross-sections (lemon yellow lines)  the thermalization radius is orders of magnitude larger than the stellar core radius ($r_{te}\gg r_{\star c}$); $r_{te}$ is even larger than the characteristic radius of the \nfw~($r_s$), although does not reaches the virial radius ($r_{vir}$). Both $r_s$ and $r_{vir}$, which are the same for the two cross-sections, are shown in  Fig.~\ref{fig:dmcoresize17} as the dotted red line and the dotted dashed red line, respectively.      
%

%
%%%%%%%%%%%%%%%%%%%%%%%%%%%%%%%%%%%%%%%%%%%%%%%%%%%%%%%%%%%%%% 
%
\section{Conclusions}\label{sec:conclusions}
\citet{2024ApJ...973L..15S} argued that the stellar cores observed in \ufd s reveal underlying \dm\ cores that cannot be created by stellar feedback on collision-less C\dm\ halos.  Assuming the validity of the claim, this work investigates the constraints imposed by such cores if they arise from self-interactions between dark matter particles (i.e., from \sidm).

The scaling relation by \citet{2023MNRAS.523.4786O} is used in Sect.~\ref{sec:sidmxsec} to work out  the \sidm\ cross-sections able to generate the cores. They are given in Eq.~(\ref{eq:ufdlimit}) for the typical velocity in Eq.~(\ref{eq:veff2}); see the red symbol with error bars in Fig.~\ref{fig:read_Ghosh+22Fig9}. The observed core sizes can be produced by \sidm\ halos that are in the core-formation phase (implying low $\sigma/m$) or in the core-collapse phase (implying high $\sigma/m$). The range of possible $\sigma/m$ values is relatively wide and aligns with those reported in the literature, summarized in Fig.~\ref{fig:read_Ghosh+22Fig9}.  Several conclusions can be drawn from the figure: since stellar feedback cannot account for the presence of cores in these systems, \ufd s provide a clean  anchor point for constraining the velocity-dependent cross-section at low velocities. This is in contrast to more luminous dwarf galaxies, where core formation may be significantly influenced by baryonic feedback. The observed scatter in cross-sections for different galaxies may arise from known but uncorrected biases (such as neglecting baryonic effects) as well as from less well-understood systematics, including variations in the definition of velocity and cross-section employed by different authors. It may also be real and reflect intrinsic diversity in galaxy properties.

The exercise carried out in Sect.~\ref{sec:sidmxsec} is complemented in  Sect.~\ref{sec:toy_model_new} with a simple model for the expected dependence on stellar mass of the stellar core radius produced by \sidm .  The advantage of modeling stellar mass and stellar core radius is that both are quantities relatively easy to observe, and should be closely related under \sidm\ scenarios. The time evolution of core \sidm\  profiles by \citet{2024JCAP...02..032Y} allows us to predict core sizes given galaxy masses and ages.  According to this model,  the stellar cores in \ufd s require cross-sections similar to the ones inferred in Sect.~\ref{sec:sidmxsec} (see Figs.~\ref{fig:dmcoresize13} and \ref{fig:dmcoresize13_3}). Moreover, the modeling reveals a trend of increasing stellar core radius with stellar mass, consistent with the behavior observed in more massive dwarf galaxies. \dm\ with other natures, like fuzzy \dm , do not comply with this observational constraint \citep[e.g.,][]{2020ApJ...904..161B,2025PDU....4902010B}. Stellar feedback qualitatively predicts the trend, but quantitatively falls too short to explain the stellar core sizes at $M_\star< 10^6\,M_\odot$  (see the gray shaded region in Fig.~\ref{fig:dmcoresize13}). We examine the sensitivity of the model predictions to variations in its free parameters, finding no significant impact on the predicted global trends. They lead to variations in stellar core radii by up to a factor of two, which are smaller than the effect of changing the parameters defining the cross-sections and also smaller than the scatter among the observed galaxies (see Fig.~\ref{fig:dmcoresize13}). The observed scatter is too large to be accounted for by the toy model, even when the uncertainties are considered. As it was mentioned above,  the origin of this  diversity (artificial or real) remains to be explained. If real, it may be due to differences in the star-formation histories followed by different galaxies \citep[e.g.,][]{PhysRevLett.119.111102,2025PhRvD.111j3041R}.    

We find that the central \dm\ densities of the \sidm\ halos in the model are consistent with the observed values, which increase with decreasing core size to yield a product $\rho_\dm (0)\, r_c$ approximately constant at  around $44\,M_\odot\,{\rm pc}^{-2}$ (see Fig.~\ref{fig:dmcoresize13_2}). The constancy  arises from the halo mass–concentration relation found in C\dm\ numerical simulations, under the assumption that the observed cores result from the redistribution of mass from an initial \nfw\ profile \citep[see][and references therein]{2025Galax..13....6S}. Consequently, any core model that originates from a \nfw\ profile consistent with the mass–concentration relation and reproduces the observed core size will naturally match the observed central dark matter density. The model set up in Sect.~\ref{sec:toy_model_new}  is of this kind, which explains why it successfully recovers the central densities.

As we mentioned above, cross-sections consistent with the observed cores can be found both for \sidm\ halos in the process of forming a core, and halos in the core-collapse phase. Even if both solutions are possible, the physics giving rise to the \sidm\ distribution in the halo is different.  Only the \dm\ particles forming the core have interacted during the core-formation phase whereas the rest maintain the original distribution from collision-less  C\dm\ halos. In core-collapse, however, the whole halo is in thermodynamic equilibrium.
Section~\ref{sec:thermalr} demonstrates that, if the \dm\ dark halos in \ufd s are core-collapsing, the thermalization radii needed to account for the observed cores are quite large. They are of the order of 1~kpc for galaxies with $M_\star < 10^6\,M_\odot$ (see Fig.~\ref{fig:dmcoresize17}).
Thus, if the \dm\ is self-interacting  with a large cross-section in the limit found in our work ($\sim$200\,\cmg ; Eq.~[\ref{eq:ufdlimit}]), one would expect that the self-interaction smears \dm\ substructure below kpc scales, also in massive galaxies.
Such smearing is expected to manifest in \sidm\ simulations. Unfortunately, there are very few cosmological simulations including large \sidm\ cross-sections, and the existing ones \citep{2021MNRAS.505.5327T,2022MNRAS.517.3045C,2023ApJ...958L..39N} are focused on the properties of the satellites around milky-way like galaxies rather than on the analysis of the \dm\ substructure resulting from such extreme \sidm\ properties. New tailored simulations are required to address this problem so as to accurately trace scales around 1\,kpc and smaller. Those simulations are needed to explore possible observational effects \citep[e.g., its impact on the expected galaxy--galaxy  lensing signal;][]{2025ApJ...978...38D}.
In addition, these large cross-sections may also have some impact on the matter-power spectrum that provides the initial conditions for the galaxies to form and evolve \citep[e.g.,][]{2021JCAP...05..013E,2022JCAP...07..012G}, an uncertain issue to be tackled systematically in the future \citep[e.g.,][]{2025ApJ...986..129N}. 

The appendixes contain several byproducts of our analysis. A stellar-to-halo mass relation, going seamlessly from $10^8$ to $10^{15}$\,$M_\odot$, is worked out in Appendix~\ref{sec:appb}. Appendix~\ref{sec:appa} provides a detailed  literature revision of the cross-sections proposed to explain observations going from gravitational lensing to galaxy cluster collisions, including modeling of dwarf galaxies.
Appendix~\ref{app:sidm_fit} shows how the stellar distribution observed in \ufd s is well reproduced by the cored \sidm\ halo shape employed in this work ($\rho_{\sidm}$, defined in Eq.~[\ref{eq:rho_sim}]).

%%%%%%%%%%%%%%%%%%%%%%%%%%%%%%%%%%%%%%%%%%%%%%%%%%%%%%%%%%%%%%
\begin{acknowledgements}
  I am grateful to
  Ignacio Trujillo,
  Claudio Dalla Vecchia,
  Ethan Nadler,
  Hai-Bo Yu,
  Manoj Kaplinghat,
  Pavel Mancera Piña,
  Filippo Fraternali, and
  Jorge Peñarrubia
  for their valuable comments and discussions on specific aspects of this work.
Thanks are also due to  Scott  Carlsten for providing some of the observational data used in Fig.~\ref{fig:dmcoresize13},
and to an anonymous referee for suggesting clarifications that strengthened the arguments in the paper.
The research is partly  funded through grant PID2022-136598NB-C31 (ESTALLIDOS 8) by the Spanish Ministry of Science and Innovation (MCIN/AEI/10.13039/501100011033)  and ``ERDF A way of making Europe”.
The author has been supported by the European Union through the grant ``UNDARK'' of the Widening participation and spreading excellence programe (project number 101159929).
\end{acknowledgements}

% Please note that we have included the references below in
% order to compile the document, but we ask you to:
%   - join the .bib files when you upload your source files

\bibliography{biblio_paper161.bib}{}
\bibliographystyle{aa}

%%%%%%%%%%%%%%%%%%%%%%%%%%%%%%%%%%%%%%%%%%%%%%%%%%%%%%%%%%%%%%%
% Appendices must be placed after   \end{thebibliography}
% They will be placed automatically on a new page.
%%%%%%%%%%%%%%%%%%%%%%%%%%%%%%%%%%%%%%%%%%%%%%%%%%%%%%%%%%%%%%%
\begin{appendix}
%%%%%%%%%%%%%%%%%%%%%%%%%%%%%%%%%%%%%%%%%%%%%%%%%%%%%%%%%%%%%%%
% In the PDF output, floats should be placed
% under their own appendix, not before the title, nor after the
% title of the next appendix.

% In short appendices, onecolumn floats (\figure*
% or \table*) will generate a blank page.
% To prevent this behaviour, a few examples are provided here. 

% In case you have a lot of floating objects for little text and the 
% LaTeX engine moves the floats away from their context, the command
% \FloatBarrier of the “placeins” package will empty the
% float buffer and place all stored floats in the continuity.

% If you still encounter problems with wide floats placement,
% just use the onecolumn environment throughout the appendices.
%%%%%%%%%%%%%%%%%%%%%%%%%%%%%%%%%%%%%%%%%%%%%%%%%%%%%%%%%%%%%%%

\section{Comments on the references represented in Fig.~\ref{fig:read_Ghosh+22Fig9}.}\label{sec:appa}

Figure~\ref{fig:read_Ghosh+22Fig9} was meant to collect a representative set of $\sigma/m$ values found in the literature.
The references were selected  from the $\sim$\,700 papers citing the seminal work where \citet{2008ApJ...679.1173R} set an upper limit on $\sigma/m$ from the lack of separation between \dm\ and stars in the Bullet Cluster.
The sources are quite diverse and there is no general rule on how $\sigma/m$ and $Velocity$ are inferred from the observational data. Thus, a significant part of the scatter in Fig.~\ref{fig:read_Ghosh+22Fig9} is due to the different analyses of the observational data leading to average cross-sections and  characteristic velocities. 
This appendix provides the equivalence between the label given the inset of  Fig.~\ref{fig:read_Ghosh+22Fig9} and the actual reference, and it also outlines the different meaning for \sidm\ cross-section and typical velocity employed in  different works.

\noindent - {\tt Randall+08}: \citet{2008ApJ...679.1173R} is the reference work setting an upper limit after the lack of significant separation between stars (from galaxies)  and \dm\ (from lensing) in the Bullet Cluster. It is based on comparison with numerical simulations made at the date of publication. 

\noindent - {\tt SA+24}: This work. The limits are worked out in Sect.~\ref{sec:sidmxsec},  Eqs.~(\ref{eq:ufdlimit3}) and (\ref{eq:veff2}), to represent the constraints imposed by the UFDs compiled  by \citet{2024ApJ...967...72R} after the analysis carried out by \citet{2024ApJ...973L..15S}.  %The velocity used for plotting is the maximum circular velocity and, therefore, likely an upper limit to the typical relative velocities between DM particles.

\noindent - {\tt Jiang+25}: \citet{2025arXiv250323710J}  explain the formation of Intermediate Mass Black Holes (IMBHs) from the core collapse of \sidm\ halos. They successfully reproduce the observed little red dot mass function with $\sigma_0/m\sim 30$\,\cmg , $w_0=0$\,\kms  ,  and  $w_1\sim 80$\,\kms\ (Eq.~[\ref{eq:resonance}]). % see the old notes for the details.  

\noindent - {\tt Robertson+19}: Comparing Einstein radii of \sidm\ simulated galaxy clusters with those observed in the CLASH survey,  \citet{2019MNRAS.488.3646R}  set the upper limit used in the figure.

\noindent - {\tt Kahl.+15}: Explaining with \sidm\ the separation between stars and \dm\ of a galaxy falling into a galaxy cluster \citep{2015MNRAS.449.3393M}, \citet{2015MNRAS.452L..54K} need the cross-section shown in the figure. The authors also estimate the large velocity of the \dm\ of the falling galaxy relative to the ambient medium.

\noindent - {\tt Peter+13}:  Cross-section based on comparing the shape of the lensing potentials with the X-ray isophotes of the massive elliptical NGC~720.  According to \citet{2013MNRAS.430..105P}, the observations allow for the cross-section employed in the figure. The  velocity comes from the virial mass of the galaxy.

\noindent - {\tt Read+18}:  The high central density of Draco allows \citet{2018MNRAS.481..860R} to set an upper limit on the \sidm\ 
cross-section. The velocity is taken as the observed velocity dispersion along the line-of-sight.

\noindent - {\tt Harvey+19}: In \sidm\ halos, the position of brightest cluster galaxy in a galaxy cluster may be offset with respect to the \dm\ distribution. \citet{2019MNRAS.488.1572H} use \sidm\ cosmological simulations with stellar feedback to set the upper limit to the cross-section in Fig.~\ref{fig:read_Ghosh+22Fig9}. It is based on 10 observed clusters.  The velocity comes from the dynamical mass of the clusters. %given in Table~1. It is a representative value for 10 observed clusters.

\noindent - {\tt Ekert+22}:  From simulations of galaxy clusters, \citet{2022A&A...666A..41E} work out a relation between the parameter $\alpha$ of the Einasto profile and the \sidm\ cross-section. The profile observed in 12 massive clusters is consistent with C\dm , which allows them to set the upper limit used in the figure. They also provide the typical  relative velocity of the \dm\ particles.   

\noindent- {\tt Gilman+23}: \cite{2023PhRvD.107j3008G} constrain resonant dark matter self-interactions with strong gravitational lenses. The diagnostics is based on whether or not the \dm\ halos in the lens have core-collapsed. They set a fairly large upper limit at low velocities. 

\noindent - {\tt Ebert+18}: From the comparison of \sidm\ cosmological numerical simulations with the brightest central galaxy
of A2667, \cite{2018ApJ...853..109E} set the upper limit at large relative velocity included in the figure.

\noindent - {\tt Adhi.+25}: Using the weak-lensing measurements reported by \citet{2021MNRAS.507.5758S},  cosmological N-body simulations, and semi-analytical fluid simulations, \citet{2025ApJ...983...50A} set the upper limit shown in the figure. Characteristic velocities are not given. Those in the figure come from the cluster halo masses turned into velocity through Eq.~(1) in \citet{2021A&A...655A.115F}.

\noindent- {\tt Gopika+24}: Following the singular approach of \citet{2022A&A...666A..41E}, \citet{2023PDU....4201291G} analyze the X-ray \dm\ profiles of 11 relaxed galaxy groups. They give a cross-section and upper limits for relative velocities between 200 and 700~\kms .  

\noindent - {\tt Tam+23}: \citet{2023ApJ...953..169T} compare the  radial acceleration relation observed in 20 high-mass galaxy clusters by \citet{2020ApJ...896...70T} with \sidm\ cosmological hydrodynamical simulations. They discard $\sigma/m$ larger than the limit used in the figure. No characteristic velocity is given but it was estimated from the \dm\ halo mass as we did for \citet{2025ApJ...983...50A}. 

\noindent - {\tt Andrade+22}: \citet{2022MNRAS.510...54A} analyze strongly lensed images of 8 galaxy clusters to measure their \dm\ density profiles. Using their \dm\ central density, they assign the cross-section used in the figure, which can also be understood as an upper limit. A characteristic velocity is given in the paper. 

\noindent - {\tt  Shi+21}: \cite{2021ApJ...909...20S} analyze the rotation of the irregular galaxy AGC 242019, which happens to be consistent with a cuspy \nfw\ \dm\ density profile. This information is used to set the upper limit for the cross-section used in the figure. The characteristic velocity is taken form the circular velocity in their Table~2.

\noindent - {\tt Correa 21}: \citet{2021MNRAS.503..920C} models the anti-correlation between the central \dm\  densities of dwarf spheroidal satellites of the MW (Milky Way) galaxies  and their pericenter distances observed by \citet{2019MNRAS.490..231K}. Using the gravothermal fluid formalism to model the subhalo density profiles, the correlation can be reproduced  with the values of cross-section and velocity employed in the figure.  

\noindent - {\tt Kap.+16}: Seminal paper by \citet{2016PhRvL.116d1302K}  where the \dm\ cores of objects going from dwarf galaxies to clusters of galaxies are interpreted in terms of \sidm . The contribution of baryon feedback is ignored.

\noindent - {\tt Vega+21}: \citet{2021MNRAS.500..247V} model the morphology of giant arcs in galaxy clusters using N-body and SPH simulations assuming both C\dm\ and \sidm . They conclude that it is not possible to rule out a \sidm\ cross-section as the one defining the upper limit represented in the figure. The velocities is inferred from the \dm\ halo mass with the recipe used for  \citet{2025ApJ...983...50A}.

\noindent - {\tt Sag.+21}: \citet{2021JCAP...01..024S} use both lensing signals and stellar dispersions to model  eight galaxy groups and seven massive galaxy clusters. They provide two points in our figure, one for groups and another for clusters. Characteristic velocities are estimated from halo masses as done for \citet{2025ApJ...983...50A}.

\noindent - {\tt Harvey+15}: \cite{2015Sci...347.1462H} measure the separation between the \dm\ structure and the galaxies in around 30 colliding galaxy clusters. The \dm\ is located via gravitational lensing whereas imaging and X-rays provide the stars and the gas. The lack of separation between \dm\ and stars allows them to set the upper limit shown in the figure. Velocities come from halo masses as done for \citet{2025ApJ...983...50A}.

\noindent - {\tt Bradac+08}: \cite{2008ApJ...687..959B} carry out a  work similar to the one for the Bullet cluster by  \cite{2008ApJ...679.1173R}. They analyze a recently identified massive merging galaxy cluster, MACS J0025.4-1222. Lensing provides the \dm\ distribution whereas optical data and X-ray render the distribution of stars and gas, respectively. The total mass distribution is offset from the hot X-ray-emitting gas but aligned with the distribution of galaxies. This is used to set an upper limit to the \sidm\ cross-section used in the figure. Typical velocities are provided by the authors. 

\noindent - {\tt Leung+21}: \citet{2021MNRAS.500..410L} analyze the isolated dwarf irregular galaxy  WLM. Photometric data and the HI rotation curve can be reproduced with the \sidm\ cross-section used in the figure. Its value together with the characteristic relative velocity are taken from their Fig.~8.

\noindent - {\tt Leung+20}: \citet{2020MNRAS.493..320L} compare the \dm\ core size of the Fornax dwarf galaxy finding it to be  large for C\dm\ but consisten with \sidm , with the cross-section used in the figure.  As characteristic velocity, Fig.~\ref{fig:read_Ghosh+22Fig9} uses the velocity dispersion of the stars along the line-of-sight.

\noindent - {\tt Mancera+24}: \citet{2024A&A...689A.344M} fit the rotation curve of the gas-rich ultra-diffuse dwarf  AGC~114905 assuming different models of \dm , including \sidm . Two different \sidm\ models fit equality well. Figure~\ref{fig:read_Ghosh+22Fig9} uses the cross-sections  at $v_{max}$, which follow from the relation hypothesized by \citet{2023ApJ...958L..39N}.

\noindent - {\tt Ando+25}: \citet{2025arXiv250313650A} fit the radial distribution of velocity dispersions in a number of UFDs and dSph galaxies. Assuming a number of priors for the galaxy properties, they carry out a  bayesian fit  that provides a range SIDM cross-sections that are favored or discarded. The ones corresponding to velocity independent cross-sections (bottom right panels in their Figs.~3 and 4) are represented in Fig.~\ref{fig:read_Ghosh+22Fig9}. The used {\em Velocity} is the typical observed velocity dispersion provided in the manuscript.    
% ____________________________________________________________

\section{Stellar mass -- halo mass relation}\label{sec:appb}
\begin{figure}
  \includegraphics[width=\linewidth]{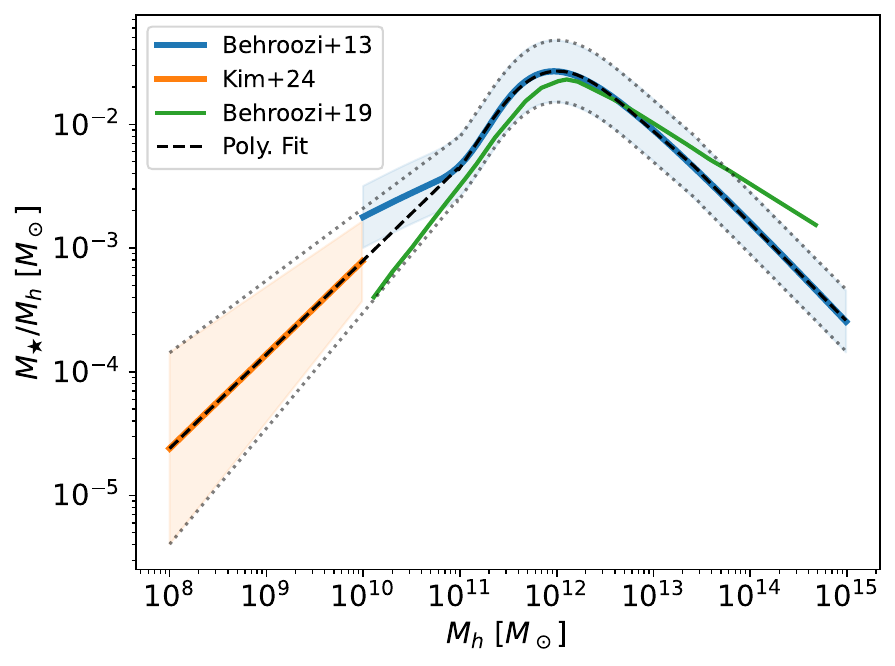}
  \caption{
Piecewise $M_\star=M_\star(M_h)$ used in Sects.~\ref{sec:toy_model_new}  and \ref{sec:thermalr} (the dashed lines; Eqs.~[\ref{eq:mass_mass}] and [\ref{eq:mass_mass2}]). A polynomial joins seamlessly the relationships worked out by \citet[][high mass]{2013ApJ...770...57B} and by \citet[][low mass]{2024arXiv240815214K}. For additional details, see Appendix~\ref{sec:appb}.
 }
\label{fig:mass_mass_relation}
\end{figure}
The analysis carried out in Sects.~\ref{sec:toy_model_new} and \ref{sec:thermalr} requires using a stellar mass  ($M_\star$)  to halo mass  ($M_h$) relation since the collisions between \dm\ particles depend on $M_h$ whereas  $M_\star$ is the true observable. For massive galaxies, $\log( M_h/M_\odot)\gtrsim 10.5$, we use the well known relation by \citet{2013ApJ...770...57B}, which is derived from observations through abundance matching. This range of masses does not include the low-mass regime of interest.  \citet{2024arXiv240815214K} provide this mass regime based on numerical simulations that goes all the way down to $\log (M_h/M_\odot)\sim 8$. Thus, we worked out a piecewise $M_\star=M_\star(M_h)$ relation coinciding with  \citet{2024arXiv240815214K}  at low mass  (the orange line in Fig.~\ref{fig:mass_mass_relation}) and with  \citet{2013ApJ...770...57B} at high mass (the blue line in Fig.~\ref{fig:mass_mass_relation}). A polynomial  seamlessly  joins  them at $\log (M_h/M_\odot)=10.976$. The full relation, shown as a black dashed line in Fig.~\ref{fig:mass_mass_relation}, is  
\begin{equation}
\log M_\star=
\begin{cases}
  -10.7 +  1.76\, \log M_h  , & \log M_h \leq 10.976,\\
\sum_{i=0}^5c_i\,x^i+\log M_h,& 10.976 \leq \log M_h \leq 13.5, \\  
8.279 + 0.209\,\log M_h, & 13.5 \leq \log M_h,
\end{cases}
\label{eq:mass_mass}
\end{equation}
with $x=(\log M_h-12.3)/1.3$, 
\begin{displaymath}
  (c_0,\dots, c_5)=
\end{displaymath}
\begin{displaymath}
  (-1.63162,-0.51845,-0.91344,0.80539,0.126597,-0.35322),
\end{displaymath}
where all the masses are  in $M_\odot$. Error bars for this function are also estimated by  joining those provided by  \cite{2024arXiv240815214K} (the orange shaded region in Fig.~\ref{fig:mass_mass_relation})  with those by \cite{2013ApJ...770...57B}  (the blue shaded region in Fig.~\ref{fig:mass_mass_relation}),  
\begin{equation}
\Delta \log M_\star=
\begin{cases}
  2.18  - 0.18\, \log M_h , & \log M_h \leq 10.976,\\
  0.25, & \log M_h > 10.976,
\end{cases}  
\label{eq:mass_mass2}
\end{equation}
where $\pm \Delta \log M_\star$ are shown as gray dashed dotted in Fig.~\ref{fig:mass_mass_relation}. 
These error bars are quite generous and include other $M_\star(M_h)$ estimates \citep[e.g.,][represented in Fig.~\ref{fig:mass_mass_relation} as a green line]{2019MNRAS.488.3143B}.

%%%%%%%%%%%%%%%%%
\section{Details of the observations represented in Fig.~\ref{fig:dmcoresize13}}\label{sec:appc}

The appendix gives the references for the observations plotted in  Fig.~\ref{fig:dmcoresize13}. It also explains how the core radii used originally were matched to the definition employed in the work  (Eq.~[\ref{foot:1}]).

\noindent - The core radii of the \ufd s from \citet{2024ApJ...967...72R} are computed from the fits by \citet{2024A&A...690A.151S} to the stellar surface density $\Sigma(R)$, and assuming that $R_{\star c}\simeq r_{\star c}$ with $\Sigma(R_{\star c})=\Sigma(0)/2$, an approximation that holds for Schuster-Plummer profiles \citep[see,][]{2022Univ....8..214S,2025Galax..13....6S}.

\noindent - The galaxy Nube, discovered by  \citet{2024A&A...681A..15M}, is particularly large for its mass. It has a conspicuously large core. The discoverers give its stellar mass, $4\times 10^8~M_\odot$, and  its projected core radius, $R_{\star c}\simeq 6.6$\, kpc. As for the \ufd s, we assume $R_{\star c}\simeq r_{\star c}$.

\noindent - The radii for the 21 classical dwarf satellites of the MW  are from the paper by \citet[][Table~1]{2025PDU....4902010B}  who collected them from the literature \citep{2008ApJ...684.1075M,2015ApJ...813..109D,2018ApJ...860...66M}. Rather than full profiles, the original references provide only effective radii, so the presence of stellar cores is taken for granted.
We transform $R_e$ to $r_{\star c}$ assuming that the central stellar distribution is well reproduced by a Schuster-Plumer profile, for which $r_{\star c}\simeq 0.565\,R_e$  \citep[see,][]{2022Univ....8..214S,2025Galax..13....6S}.

%\noindent - The massive early type galaxies with cores are from \citet{2013AJ....146..160R}. These small cores are supposed to be created by the  scouring of stars by mergers of binary supermassive black holes. Using the fits to the photometric profiles provided by the authors, we computed the core radii. Masses are inferred from absolute magnitudes assuming a mass-to-light ratio of 2, typical of old stellar populations.  

\noindent - \citet{2022ApJ...933...47C} compiled satellites of MW-like galaxies. Their stellar mass follows from integrated photometry as given by \citet{2013MNRAS.430.2715I}. The core radii are obtained from the analysis by \citet{2024A&A...690A.151S}, where the observed radial surface density profiles are fit using projected polytropes. Polytropes always have cores \citep[e.g.,][]{2020A&A...642L..14S}. From these fits, one directly recovers $r_{\star c}$ as defined in Eq.~(\ref{foot:1}).

\noindent - The  dSph galaxy Fornax is represented with the core radius derived from a Plummer profile fit to the observed surface density by \citet{2006A&A...459..423B}.  The stellar mass comes from \citet{2016A&A...590A..35D}. 

\noindent - The predictions from stellar feedback come from \citet[][Fig.7, right panel]{2020MNRAS.497.2393L}.  Fig.~\ref{fig:dmcoresize13} shows the analytic curve they provide $\pm$ a factor of 2 as suggested by the scatter in their Fig.~7.  These authors use a definition of $r_c$ based on core-Einasto profile fits to the numerical profiles. The Fig.~4 in their paper shows that this definition approximately agrees with ours. We note that the stellar feedback sub-grid physic implemented in the used numerical simulation \citep[FIRE2;][]{2018MNRAS.480..800H}  is particularly strong \citep[e.g.,][]{2020MNRAS.494.3971M,2021MNRAS.508.2979P} so that other simulations will likely give even smaller stellar cores.
% I asked the DM group of the IAC (including Arianna and students) for other references, but there seems to be none.

\section{Fit to the observed stellar surface density using projected \sidm\ profiles}\label{app:sidm_fit}

The stellar distribution of the six \ufd s analyzed in this work follows a well defined shape shared by all of them; see the symbols in Fig.~\ref{fig:dmcoresize15}. This shape is well reproduced by the line-of-sight projection of the \sidm\ dark matter profile assumed by \citet{2024JCAP...02..032Y} and used in Sect.~\ref{sec:toy_model_new}. The solid magenta line represents the least-squares best fit of the observed points using the profile in Eq.~(\ref{eq:rho_sim}) projected along the line-of-sight to generate surface densities. The fit uses as free parameters $\rho_s'$, $r_s'$, and $r_c'$. The best fit yields $r_c'\simeq 1.1\,r_s'$. 
\begin{figure}
\includegraphics[width = 0.9\linewidth]{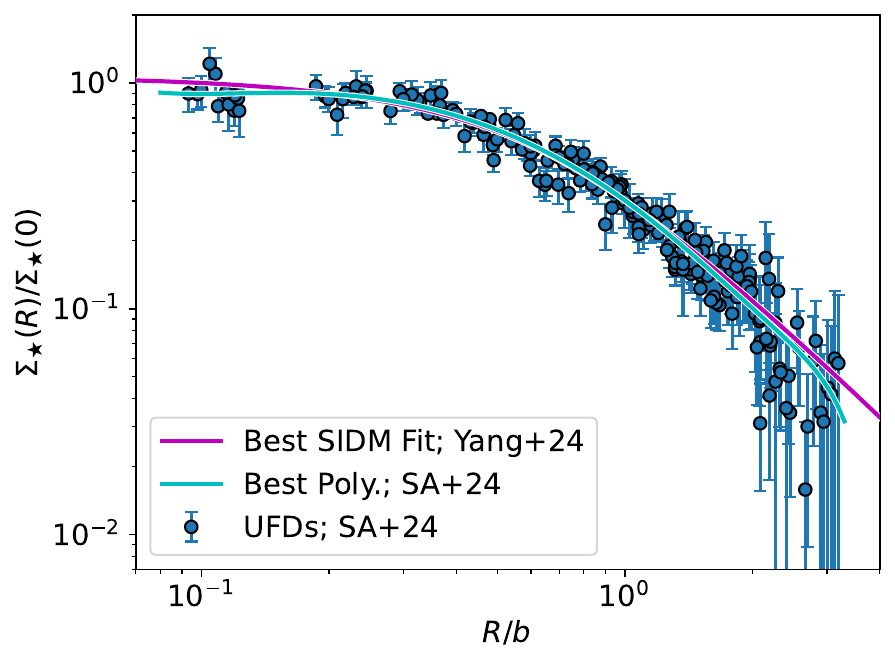}
\caption{Observed stellar surface density distribution for the six \ufd s analyzed in this work as worked out in \citet[][their Fig.~1]{2024ApJ...973L..15S} (the symbols with error bars). It is well reproduced by the line-of-sight projection of the \sidm\ dark matter profile  used in Sect.~\ref{sec:toy_model_new}. The magenta line corresponds to the best fit using one of such profiles. The figure also includes the polynomial fit to $\Sigma_\star(R)$ given by \citet[][the solid cyan line]{2024ApJ...973L..15S}. The symbol $R$ stands for the projected radial distance from the center of the galaxy, whereas $b$ is a radial scaling factor, different for the different dwarfs, that allows the whole set of dwarfs to collapse to a single shape.}
\label{fig:dmcoresize15}
\end{figure}

\end{appendix}
\end{nolinenumbers}
\end{document}